\documentclass[reqno]{amsart}
\usepackage{amsmath}
\usepackage{amssymb}
\usepackage[T1]{fontenc}
\usepackage{lmodern}  
\usepackage[utf8]{inputenc}
\usepackage{graphicx}
\usepackage{float} 
\usepackage{color}
\usepackage{comment}

{ 

\begin{document}

\title{Non-equilibrium theories of rarefied gases: internal variables and extended thermodynamics 
}



\author{R. Kov\'acs$^{123}$, D. Madjarevi\'c$^{4}$, S. Simi\'c$^{5}$ and P. V\'an$^{213}$}

\address{
$^1$Department of Energy Engineering, Faculty of Mechanical Engineering, BME, Budapest, Hungary and
$^2$Department of Theoretical Physics, Institute for Particle and Nuclear Physics, Wigner Research Centre for Physics, Budapest, Hungary and
$^3$Montavid Thermodynamic Research Group and
$^{4}$ Department of Mechanics, Faculty of Technical Sciences, University of Novi Sad, Novi Sad, Serbia, and
$^5$ Department of Mathematics and Informatics, Faculty of Sciences, University of Novi Sad, Novi Sad, Serbia}
\date{\today}




\begin{abstract}
Limits of classical constitutive laws such as Fourier and Navier-Stokes equations are discovered since decades. However, the proper extensions -- generalizations of these are not unique. They differ in the underlying physical principles and in modeling capabilities. In this paper, two different theories are discussed and compared to each other, namely the kinetic theory-based Rational Extended Thermodynamics (RET) and non-equilibrium thermodynamics with internal variables (NET-IV).

First, the paper starts with the case of rigid heat conductors summarizing the result achieved so far. Then a typical example of compressible bodies is shown by presenting the first generalization for rarefied gases, called Meixner's theory. It is further extended using generalized entropy current in the framework of NET-IV. It is shown how its structure is related to RET and how the compatibility between them can be acquired.
\end{abstract}

\maketitle

\section{Motivation and preparation}
\label{intro}
The simplest and most widely applied classical continuum theory is related to isotropic fluids. They are based upon conservation laws of mass, momentum, and energy, adjoined by appropriate constitutive equations for the heat flux $q_i$, deviatoric $\Pi_{\langle ij \rangle}$ and spherical $\Pi^s$ part of the pressure:
\begin{align} \label{eq_ClassicalConstRel}
q_i &= - \lambda \partial_i T,
\nonumber \\
\Pi_{\langle ij \rangle} &=- \nu \partial_{\langle i} v_{j \rangle},
\\
\Pi^s &=- \mu \partial_i v_i,
\nonumber
\end{align}
where $T$ is the temperature, $v_i$ is velocity field, $\lambda$ is thermal conductivity, $\nu$ is shear viscosity and $\mu$ is bulk viscosity. Here $\partial_i$ is the spatial derivative, the summation convention of Einstein is used and ${\langle \; \rangle}$ denotes deviatoric part of the symmetric tensor. These equations are known as Fourier \eqref{eq_ClassicalConstRel}$_{1}$ and Navier-Stokes \eqref{eq_ClassicalConstRel}$_{2-3}$ constitutive relations.

Although widely accepted and applied in a multitude of problems, there are numerous thermomechanical phenomena that constitutive relations \eqref{eq_ClassicalConstRel} are not capable of describing properly. In the sense of modeling, either linear relations between the constitutive quantities and gradients of field variables are inappropriate, or some kind of inertia of constitutive quantities has to be taken into account. Shortcomings of the classical model especially appear in processes under extreme thermomechanical conditions, or when the field variables are not restricted to a small neighborhood of local equilibrium state. These difficulties could be overcome in different ways. Here, in order to motivate our study, an example of modeling of the heat waves in solids will be presented.

\subsection{Motivating example: the heat waves}
\label{subs:1}

In the case of heat conduction, the wave-like propagation of heat appeared as a non-Fourier phenomenon. It was first measured by Peshkov in superfluid He$^4$ \cite{Pesh44}. Later on, the ballistic heat conduction required the next level of modeling which still remains to be a theoretical challenge  \cite{FriCim95,FriCim96,FriCim98,KovVan16}. In current perspective, promising modeling frameworks, and at the same time thermodynamically consistent, are based upon Rational Extended Thermodynamics (RET) \cite{MulRug98b,DreStr93a} and non-equilibrium thermodynamics with internal variables and current multipliers (NET-IV) \cite{KovVan15,KovVan18,BerVan17b}. Their application to the heat wave modeling will be briefly described in the sequel.

In RET, two different approaches to phonon gas can be taken to capture the propagating (wave) character of heat. The first one is continuum one, exposed in detail in \cite{RuggMurSecc94}, that describes the heat propagation using the energy conservation law and the balance law for the heat flux:
\begin{align} \label{eq_RuggSecondSound}
  \rho \dot{e} + \partial_i q_i & = 0,
  \\
  \frac{d}{dt}(\alpha q_i) + \partial_i \nu & = - \frac{\nu'}{\kappa} q_i,
  \nonumber
\end{align}

\noindent where dot indicates the material derivative and prime indicates the derivative with respect to temperature $T$; $\kappa(T)$ is the heat conductivity and $\alpha(T)$ and $\nu(T)$ are constitutive functions related to the second sound. The system of governing equations is closed by means of the compatibility with the entropy principle. The advantage of this model lies in its nonlinearity and wave speeds that depend on temperature. This enables the explanation of the shape changes of the second sound profile. For $\alpha = \mathrm{const.}$, governing equations \eqref{eq_RuggSecondSound} can be reduced to Cattaneo equation \cite{Cattaneo58} which predicts the constant wave speed.

Another approach within RET is more closely related to the kinetic theory of gases, and phonon gas equations are obtained as proper moment equations of the Boltzmann-like equation for the phonon phase density function \cite{DreStr93a}. Closure of the governing equations, although trivial in this case, is based upon the maximization of entropy. According to this approach, one has to define the way how phonons interact with each other. Thus, two particularly interesting processes are introduced. One of them is the so-called normal process, which leads to wave-like propagation, the Maxwell-Cattaneo-Vernotte (MCV) \cite{Max1867,Cattaneo58,Vernotte58,JosPre89,JosPre90a} equation is obtained in this case. The other one is the resistive process resulting in diffusive propagation where the Fourier equation holds. However, during ballistic heat conduction, there is no interaction between phonons, it is a propagation without any interaction among them, the phonons are scattered only on the boundary. The resulting system of moment equations in the one-dimensional case reads as \cite{MulRug98b,DreStr93a}:
\begin{align}
\partial_t e + c^2 \partial_x p &= 0, \nonumber \\
\partial_t p + \frac{1}{3} \partial_x e +\partial_x N &= -\frac{1}{\tau_R} p, \label{eq_kt_ball} \\
\partial_t N + \frac{4}{15} c^2 \partial_x p &= - \left( \frac{1}{\tau_R}+\frac{1}{\tau_N} \right ) N, \nonumber
\end{align}
where $e$, $p$ and $N$ are the moments, i.e. the energy density, momentum density, and the deviatoric part of the pressure tensor. The relaxation times $\tau_R$ and $\tau_N$ are related to the interacting processes. The other coefficients in the system (\ref{eq_kt_ball}) are calculated according to the molecular structure, only the relaxation times are to be fitted for experiments.

In NET-IV, an analogous material model, called ballistic-conductive equation, can be derived for heat conduction using an extended entropy density and a generalized entropy current density with internal variables $\xi_i$ and $Q_{ij}$ \cite{KovVan15,FamEta18m}:
\begin{align}
s(e, \xi_i, Q_{ij}) = s_{eq} (e) - \frac{m_1}{2} \xi_i \xi_i - \frac{m_2}{2} Q_{ij} Q_{ij}, \quad J_i = b_{ij} \xi_j, \label{NETIVeq4}
\end{align}
where $s_{eq}$ corresponds to term of local equilibrium, $m_1$ and $m_2$ are positive parameters. These functional assumptions characterize the deviation from local equilibrium according to thermodynamic requirements. There are two internal variables in the theory, a vectorial, $\xi_i$, and a tensorial $Q_{ij}$. In the entropy current $J_i$, the $b_{ij}$ is a current multiplier \cite{Nyiri91} allowing coupling between the evolution equations of different tensorial order quantities in isotropic materials, too. This coupling stands as a requirement to appropriately describe ballistic effects \cite{KovVan16}.  Later on the structure of the equations show that the interpretation of the internal variables is natural -- $\xi_i$ is connected to the heat flux $q_i$  \cite{VanFul12a} and $Q_{ij}$ is the flux of the heat flux, the dissipative pressure. One may chose to fix the interpretation of the internal variables from the beginning. E.g. the vectorial internal variable was chosen as heat flux in \cite{KovVan15}, however it is not necessary, see \cite{VanFul12a}. All internal variables are fixed as dissipative fluxes in Extended Irreversible Thermodynamics. The NET-IV approach results in a compatible system in linearized one-dimensional form (with identification $\xi = q$):
\begin{align}
\rho \partial_t e + \partial_x q &=0, \nonumber \\
\tau_q \partial_t q +  q + \lambda \partial_x T + \kappa \partial_x Q &= 0, \label{bc_eq3} \\
\tau_Q \partial_t Q +  Q + \kappa \partial_x q &= 0, \nonumber
\end{align}
containing the same terms as \eqref{eq_kt_ball}, but with different coefficients. Although the coefficients can be compared to each other, and one can make a correspondence between them, it remains arbitrary since the coefficients in the NET-IV model \eqref{bc_eq3} are not fixed as in RET model \eqref{eq_RuggSecondSound}. There is more freedom for fitting with experimental data, but at the same time, one has to deal with more parameters. This leads to another distinction between these two approaches in the case of ballistic propagation: the NET-IV model, with properly fitted parameters, may recover the exact propagation speed, whereas the RET model requires at least 30 moments equations for an accurate approximation. The difference in the interpretation of the fields is also remarkable. We will see that in RET, the starting points are the evolution equations in a detailed balance form and then the entropy density and entropy flux are derived as well as the entropy production. In NET-IV, the starting point is the entropy balance, and then the evolution equations are derived.

It is worth mentioning here the role of Extended Irreversible Thermodynamics (EIT), which can be considered as a bridge between RET and NET-IV \cite{CimEta14a}. There the dissipative fluxes are the independent variables, in particular the heat flux. The form of their evolution equations is postulated, they are balances in a hierarchical structure, like in RET \cite{JouVasLeb88ext,JouEtal10b,AlvEta09a}. That assumption is applied as an extension of entropy density, similarly to Eq.~(\ref{NETIVeq4})$_{1}$. Then the entropy production is calculated and used to close the system of evolution equations. Moreover, in recent extensions of the original EIT theory, the entropy current density is a-priori generalized, also using the gradient of fluxes \cite{JouEtal10b,SellEtal16b,SellEtal13}. It is compatible with both RET and NET-IV. EIT is indeed a transition from RET to NET-IV: it keeps some assumptions from kinetic theory but still offers flexibility regarding the coefficients \cite{SellEtal16b}.

All the models of heat waves presented here were tested against the experimental results. Continuum model \eqref{eq_RuggSecondSound} was used to determine the critical temperature at which the second sound in NaF and Bi was identified experimentally \cite{RuggMurSecc90}. Kinetic-theory-based model \eqref{eq_kt_ball} was used for numerical simulations that recover three different modes---the ballistic phonons, the second sound, and the diffusive regime \cite{MulRug98b}. Finally the NET-IV model \eqref{bc_eq3} is successfully tested on reconstructing the NaF experiments \cite{KovVan18}.

\subsection{Aim and outline of the paper}
\label{subs:2}

The kinetic picture of ballistic propagation may indicate the same behavior in rarefied gases, too. To that end, the present study has two principal aims: first, to exploit the NET-IV framework to develop generalized Navier-Stokes-Fourier for rarefied gases; second, to perform a thorough comparison with RET as a paradigmatic non-equilibrium modeling framework for rarefied gases. The first aim of the study comes out as an attempt to apply to rarefied gases a well-established method for modeling non-equilibrium processes. Indeed, the method of internal variables was applied to a multitude of phenomena that may be regarded as non-equilibrium ones (see the book of Maugin \cite{Maugin} for a nice review), but there was no systematic analysis of rarefied polyatomic gases in this framework. Once this goal is reached, there naturally appears the second one---a comparison with RET of polyatomic rarefied gases. Both of them are thermodynamically consistent and aim at the proper description of non-equilibrium phenomena. Nevertheless, each one of them has its own peculiarities regarding aims, concepts, and methodology, which calls for careful comparative analysis.

The analogy of ballistic heat conduction and real gases is in the type of particles: in rarefied gases instead of phonons, we have real molecules like normal hydrogen. This fact is exploited in the framework of NET-IV when the coupled system of Fourier and Navier-Stokes equations is generalized. That is, the same form of entropy density and entropy current is applied, but the mass density $\rho$ is considered as a variable now. The goal of the extension of classical material laws is to study the irreversible processes that are coupled to the propagation of sound, especially close to a shock wave in rarefied gases. To that end, RET of polyatomic gases \cite{RETpoly} and NET-IV \cite{BerVan17b,KovVan15} seem to be proper counterparts aimed to describe the same class of physical phenomena.

In this work, we compare these two theories with the aim of finding a common theoretical background for them. This goal is going to be reached by seeking the answer to the following questions:
\begin{enumerate}
  \item How do we construct the system of governing (balance) equations?
  \item How do we close the governing system?
  \item How do we determine the material parameters (constitutive quantities)?
  \item Under what circumstances we obtain the equivalent systems?
\end{enumerate}
Although formal in their nature, these questions will help us to reveal the common ground and better understand the conceptual subtleties of the two approaches. To that end, the paper will be organized as follows. In Section \ref{subs:3}, the comparative analysis of the Meixner theory, the first extension of the Navier-Stokes-Fourier (NSF) system, will be performed from the point of view of NET-IV and RET. Then, in Section \ref{sec:NET-IV}, we move forward using the generalized entropy current density, together with the internal variables, to derive the generalized Navier-Stokes-Fourier equations. Subsequent Section \ref{sec:14moments}, the 14 moments RET model of polyatomic gases is presented in the form amenable to comparison with the NET-IV approach given in Section \ref{sec:NET-IV}. A discussion about the similarities and discrepancies of two approaches in the sense of governing equations, concepts, and procedures is given in Section \ref{sec:Comparison}.

In this Section, we shall make the first step towards the comparison of different theories of rarefied gases. Our aim is to compare the Meixner model, the simplest model of non-equilibrium processes that can be derived within the framework of internal variables theory, with the extended thermodynamic model with 6 fields (ET6), which takes into account only the dynamic pressure as a non-equilibrium variable.

\subsection{Internal variables---the Meixner model}
\label{subs:3}

In the following, the Meixner model with a single internal variable will be presented. An appropriate background can be found in \cite{Meix43a,GrooMaz63non}. The starting point is the usual set of conservation laws of mass, momentum, and energy:
\begin{align} \label{IV-CLaws}
  \dot \rho + \rho \partial_i v_i & = 0, \nonumber \\
  \rho \dot v_i + \partial_j P_{ij} & = 0_i, \\
  \rho \dot e + \partial_i q_i + P_{ij} \partial_i v_j & = 0. \nonumber
\end{align}
In \eqref{IV-CLaws} $e$ is the specific internal energy, $P_{ij}$ is the pressure tensor and $q_{i}$ is the heat flux. $P_{ij}$ and $q_{i}$ can be regarded as non-convective (conductive) parts of momentum and energy flux, respectively. We shall assume that pressure tensor has the following representation:
\begin{equation}\label{IV-PressureTensor}
  P_{ij} = \Pi_{ij} + p \delta_{ij}; \quad
  \Pi_{ij} = \Pi_{\langle ij \rangle} + \Pi \delta_{ij}, \; \Pi = \frac{1}{3} \Pi_{kk},
\end{equation}
where $p$ is the hydrostatic pressure, $\Pi_{\langle ij \rangle}$ is the traceless deviatoric part of the viscous stress tensor and $\Pi$ is the dynamic pressure ($\delta_{ij}$ is the Kronecker symbol).

Meixner's theory characterizes the deviation from local equilibrium by a single scalar internal variable. The entropy density is additively decomposed into an equilibrium part and an internal variable dependent non-equilibrium part:
\begin{equation} \label{IV-MeixnerEntropy}
s(e,\rho,\xi) = s_{eq}(e,\rho) - \frac{m_1}{2}\xi^{2}.
\end{equation}
It is assumed that equilibrium entropy density obeys the classical Gibbs relation:
\begin{equation}\label{IV-Gibbs}
  T ds_{eq} = de - \frac{p}{\rho^{2}} d\rho.
\end{equation}
Crucial step in the analysis is exploitation of the entropy balance law, $\rho \dot{s} + \partial_{i} J_{i} = \sigma_{s} \geq 0$, with classical entropy current $J_i = q_i / T$:
\begin{equation*}
\sigma_s = \rho \dot s + \partial_i J_i
  = \rho \left( \dot{s}_{eq} - m_{1} \xi \dot{\xi} \right)
    + q_i \partial_i \left( \frac{1}{T} \right) + \frac{1}{T} \partial_i q_i.
\end{equation*}
Taking into account Gibbs relation \eqref{IV-Gibbs} and conservation laws \eqref{IV-CLaws}$_{1}$ and \eqref{IV-CLaws}$_{3}$, after straightforward calculation one arrives at:
\begin{equation}
  \sigma_{s} = - \frac{1}{T} \Pi_{ij} \partial_i v_j
    + q_i \partial_{i} \left( \frac{1}{T} \right) - \rho m_1 \xi \dot{\xi} \geq 0.
\end{equation}
The final form of the entropy production rate $\sigma_{s}$ is obtained when the viscous stresses are decomposed as in \eqref{IV-PressureTensor}, as well as the velocity gradient, $\partial_{i} v_{j} = \partial_{\langle i} v_{j \rangle} + \partial_{k} v_{k} \delta_{ij}$:
\begin{equation} \label{IV-MeixnerInequality}
  \sigma_{s} = - \frac{1}{T} \Pi_{\langle ij \rangle} \partial_{\langle i} v_{j \rangle}
    - \frac{1}{T} \Pi \partial_{k} v_{k}
    + q_i \partial_{i} \left( \frac{1}{T} \right) - \rho m_1 \xi \dot{\xi} \geq 0.
\end{equation}
Following the standard procedure, thermodynamic fluxes and forces are introduced \cite{Onsager31I,Onsager31II,Verhas97}, given in Table \ref{IV-TableFluxForce}:
\begin{table}[h!]
\centering
\begin{tabular}{c|c|c}
  & Fluxes & Forces \\ \hline
  Fluid -- traceless & $\Pi_{\langle i j \rangle }$ &
    $- \frac{1}{T} \partial_{\langle i} v_{j \rangle}$ \\ \hline
  Fluid -- spherical & $\Pi$ &
    $- \frac{1}{T} \partial_{k} v_{k}$ \\ \hline
  Thermal & $q_{i}$ &
    $ \partial_i \left( \frac{1}{T} \right)$ \\ \hline
  Internal & $\rho \dot{\xi}$ & $- m_1 \xi$ \\
\end{tabular}\\
\caption{Thermodynamic fluxes and forces}
\label{IV-TableFluxForce}
\end{table}
The simplest way to guarantee the non-negative entropy production rate $\sigma_{s}$ in every thermodynamic process is to establish proper linear relations between forces and fluxes, taking into account possible cross-effects between the terms of the same tensorial order:
\begin{align}\label{IV-ConstRelations}
  \Pi_{\langle ij \rangle} & = - \frac{\mu}{T} \partial_{\langle i} v_{j \rangle}, \nonumber \\
  q_{i} & = \lambda \partial_i \left( \frac{1}{T} \right), \\
  \rho \dot{\xi} & = - l_{11} m_{1} \xi - \frac{l_{12}}{T} \partial_{k} v_{k}, \nonumber \\
  \Pi & = - l_{21} m_{1} \xi - \frac{l_{22}}{T} \partial_{k} v_{k}. \nonumber
\end{align}
First two relations represent the classical constitutive relations of Navier-Stokes-Fourier theory, the third one determines the evolution of the internal variable and the last is a constitutive relation for the dynamic pressure. Also, from \eqref{IV-ConstRelations}$_{4}$ one may observe the physical interpretation of internal variable $\xi$---it is a linear combination of dynamic pressure $\Pi$ and divergence of velocity (divided by the temperature). In the special case, $l_{22} = 0$, we have $\Pi = - l_{21} m_{1} \xi$ and $\xi$ is directly related to dynamic pressure.

Due to the scalar character of internal variable and linear relations between forces and fluxes in \eqref{IV-ConstRelations}$_{3-4}$, it is possible to eliminate $\xi$ and obtain a single constitutive relation of the rate type for dynamic pressure $\Pi$. Therefore, the final set of equations for rarefied gases, obtained by means of the single internal variable, reads:
\begin{align} \label{IV-BLaws}
  \dot \rho + \rho \partial_i v_i & = 0, \nonumber \\
  \rho \dot v_i + \partial_j P_{ij} & = 0_i, \nonumber \\
  \rho \dot e + \partial_i q_i + P_{ij} \partial_i v_j & = 0 \\
  \rho \dot{\Pi} + l_{11} m_{1} \Pi & =
    - (l_{11} l_{22} - l_{12} l_{21}) m_{1} \frac{\partial_{k} v_{k}}{T}
    - l_{22} \rho \frac{d}{dt} \left( \frac{\partial_{k} v_{k}}{T} \right) \nonumber \\
  \Pi_{\langle ij \rangle} & = - \frac{\mu}{T} \partial_{\langle i} v_{j \rangle}, \nonumber \\
  q_{i} & = \lambda \partial_i \left( \frac{1}{T} \right), \nonumber
\end{align}
where \eqref{IV-PressureTensor} has to be taken into account.

\subsection{Extended thermodynamics with 6 fields}
\label{subs:4}
Within the realm of RET of polyatomic gases \cite{RETpoly}, the theory with 6 fields (ET6) is the simplest possible theory that captures non-equilibrium effects \cite{AriEtal12c}. It contains only one non-equilibrium variable---the dynamic pressure $\Pi$. Its development was fostered by two facts: first, experimental evidence shows that in polyatomic gases, dynamic pressure has a greater influence than shear stresses and heat flux (see \cite{Tanietal14b}); second,  it can be easily compared with the simplest version of Meixner's theory \cite{Arietal15}. In this Section, it will be presented only in the simplest possible form, without the details of its derivation.

The governing equations of ET6 have the following form \cite{AriEtal12c}:
\begin{align}\label{ET6-BLaws}
  & \dot \rho + \rho \partial_k v_k = 0, \nonumber \\
  & \rho \dot v_i + \partial_{i} (p + \Pi) = 0_i, \\
  & \rho \dot e + (p + \Pi) \partial_k v_k = 0 \nonumber \\
  & \frac{d}{dt} \left[ \frac{p + \Pi}{\rho} - \frac{2}{3} e \right]
    = - \frac{1}{\tau} \frac{\Pi}{\rho}, \nonumber
\end{align}
where we assumed the linearized form of the source term in \eqref{ET6-BLaws}$_{4}$. In general case, ET6 theory is a peculiar extended theory which admits the possibility of non-linear closure \cite{Arietal15}. However, for the purpose of comparison with internal variables approach it is sufficient to keep only the linear approximation. Entropy density and entropy flux are:
\begin{equation}\label{ET6-Entropy-EntropyFlux}
  s = s_{eq} - \frac{\Psi(\rho,e)}{\rho} \Pi^{2}, \quad
  J_{i} = 0_{i},
\end{equation}
whereas the entropy balance law reads:
\begin{equation}\label{ET6-EntropyBalance}
  \frac{d}{dt} \left[ s_{eq} - \frac{\Psi(\rho,e)}{\rho} \Pi^{2} \right]
    = \frac{2 \Psi}{\rho \tau} \Pi^{2}.
\end{equation}

\subsection{Comparison of IV with ET6}
\label{subs:5}
This part of our study aims to compare the equations \eqref{IV-BLaws} of the internal variable approach with ET6 equations \eqref{ET6-BLaws}. It is obvious that the model \eqref{IV-BLaws} contains more field variables than \eqref{ET6-BLaws}. It consists of conservation laws of mass, momentum, and energy, and constitutive relations for dynamic pressure $\Pi$, pressure deviator $\Pi_{\langle ij \rangle}$ and heat flux $q_{i}$. The only non-classical part of the model is equation \eqref{IV-BLaws}$_{4}$. It came out as a consequence of the second law of thermodynamics but has the form of the balance law rather than classical constitutive relation. On the other hand, ET6 model comprises the same conservation laws, but only one additional equation---balance law for dynamic pressure $\Pi$. Pressure deviator and heat flux are neglected. Moreover, the formal procedure of RET assumes from the outset the form of balance laws for all the state variables---$\Pi$ belongs to this set. Compatibility with entropy inequality yields the form of the unknown non-convective fluxes and the source terms. This will be thoroughly discussed in Section \ref{sec:14moments}.

At this level of approximation, the internal variable approach yields the system, which is inherently parabolic. Not only because of the constitutive relations \eqref{IV-BLaws}$_{5-6}$, but the balance law \eqref{IV-BLaws}$_{4}$ as well, since it contains the material derivative of $\partial_{k} v_{k}/T$ which brings second order derivatives. On the other hand, ET6 model \eqref{ET6-BLaws} is hyperbolic by construction.

To be comparable with ET6 model, it is necessary to reduce the number of field variables in the internal variable model \eqref{IV-BLaws} by neglecting the pressure deviator, i.e. $P_{ij} = (p + \Pi) \delta_{ij}$, $\Pi = (1/3) \Pi_{ii}$, and the heat flux by taking $\mu = 0$, $\lambda = 0$. Also one must assume $l_{22} = 0$, which altogether leads to the following reduced system:
\begin{align}\label{IV-BLawsReduced}
  & \dot \rho + \rho \partial_i v_i = 0, \nonumber \\
  & \rho \dot v_i + \partial_{i} (p + \Pi) = 0_i, \\
  & \rho \dot e + (p + \Pi) \partial_i v_i = 0 \nonumber \\
  & \frac{\rho}{l_{11} m_{1}} \dot \Pi + \Pi =
    \frac{1}{T} \frac{l_{12} l_{21}}{l_{11}} \partial_k v_k. \nonumber
\end{align}
On the other hand, equation \eqref{ET6-BLaws}$_{4}$ has to be adapted by using the ideal gas equations of state, $e = p/\rho(\gamma - 1)$, and taking into account the conservation laws for mass and energy. After certain straightforward calculations, the resulting system reads:
\begin{align}\label{ET6-BLawsReduced}
  & \dot \rho + \rho \partial_k v_k = 0, \nonumber \\
  & \rho \dot v_i + \partial_{i} (p + \Pi) = 0_i, \\
  & \rho \dot e + (p + \Pi) \partial_k v_k = 0 \nonumber \\
  & \tau \dot \Pi + \tau \left[ \Pi - \left( \gamma - \frac{5}{3} \right) (p + \Pi)
    \right] \partial_{k} v_{k} = - \Pi. \nonumber
\end{align}
Comparison of the reduced IV-Meixner model \eqref{IV-BLawsReduced} with ET6 model \eqref{ET6-BLawsReduced} yields the following equivalence conditions:
\begin{equation}\label{IV-ET6_Equivalence}
  \tau_{\Pi} := \frac{\rho}{l_{11} m_{1}} = \tau; \quad
  \frac{1}{T} \frac{l_{12} l_{21}}{l_{11}} = \tau \left[ \Pi - \left( \gamma - \frac{5}{3}
    \right) (p + \Pi) \right]; \quad \mu = 0; \quad \lambda = 0.
\end{equation}

The compatibility is complete for the full nonlinear set of equations, and it differs from the one given in \cite{Arietal15}, because of the different identifications of the internal variable and static pressure.

Further comparison of IV and RET approach to rarefied gases will be postponed until more refined theories are developed in the next two Sections.

\section{NET-IV: generalized Navier-Stokes-Fourier equations}
\label{sec:NET-IV}

As it is demonstrated in Section \ref{subs:3}, the Meixner model obtained by extension of entropy density itself is of limited capacity for capturing non-equilibrium effects---it does not obtain coupling between the heat flux and the pressure tensor. To get a proper model within the framework of internal variables, which inherits non-equilibrium coupling of different effects in rarefied gases, certain generalizations are needed. The Ny\'{i}ri-multipliers \cite{Nyiri91} open further possibilities within a generalized entropy current. Let us choose the same set of extensions of the state space together with the identifications $\xi_i=q_i$ and $Q_{ij} = \Pi_{ij}$, that is the vectorial internal variable is identified as the heat flux and the tensorial one as the viscous pressure. This is the usual assumption in theories of Extended Thermodynamics, a consequence of the compatibility with kinetic theory \cite{MulRug98b,JouVasLeb88ext}. Then the most general form of the entropy density for isotropic materials is
\begin{equation} \label{NET-IV-Entropy}
s(e, \rho, q_i, \Pi_{ij})=s_e(e, \rho) - \frac{m_1}{2} q_i q_i -\frac{m_2}{2} \Pi_{\langle ij \rangle} \Pi_{\langle ij \rangle} - \frac{m_3}{6} \Pi_{ii} \Pi_{jj},
\end{equation}
where the spherical and deviatoric parts are distinguished, $m_1$, $m_2$ and $m_3$ are strictly positive parameters\footnote{This is not a Taylor series expansion, but a consequence of Morse lemma, as long as the internal variables are abstract fields that characterize the deviation from local equilibrium \cite{Verhas97}}. The entropy current is generalized as:
\begin{equation}
J_i = (b_{\langle i j \rangle} + b_{kk}\delta_{ij}/3)q_j,
\label{JC}
\end{equation}
where the current multiplier, $b_{ij} = b_{\langle i j \rangle} + b_{kk}\delta_{ij}/3$,  is split into deviatoric and spherical parts, too.  Then one can calculate the entropy production, considering the usual Gibbs relation for the equilibrium specific entropy, $de=Tds+\frac{p}{\rho^2}d\rho$:
\begin{align} \label{eq:IV-EntropyProd}
\sigma_s &= \rho \dot s + \partial_i J_i = \nonumber \\
&= \rho \left ( \partial_e s \dot e + \partial_{\rho} s \dot \rho + \partial_{q_i} s \dot q_i + \partial_{\Pi_{\langle ij \rangle}}s \dot \Pi_{\langle ij \rangle} + \partial_{\Pi_{ii}}s \dot \Pi_{ii} \right ) +\partial_i \big [ ( b_{\langle ij \rangle} +\frac{1}{3} b_{kk}\delta_{ij})q_j \big ] = \nonumber \\
&=\partial_i q_i \left (\frac{b_{kk}}{3} - \frac{1}{T} \right) + q_i \left (\partial_i\frac{b_{kk}}{3} + \partial_k b_{\langle ki \rangle}- \rho m_1 \dot q_i \right)+ b_{\langle ij \rangle} \partial_i q_j -\nonumber \\ &
- \frac{\Pi_{ii} }{3}\left ( \frac{1}{T} \partial_j v_j + \rho m_3 \dot \Pi_{jj} \right) - \Pi_{\langle i j \rangle }\left ( \frac{1}{T} \partial_i v_j + \rho m_2 \dot \Pi_{\langle i j \rangle } \right) \geq 0.
\end{align}
In \eqref{eq:IV-EntropyProd} one may recognize the products of thermodynamic fluxes and thermodynamic forces, listed in Table \ref{table:fandf}:
\begin{table}[h!]
\centering
\begin{tabular}{c|c|c}
  & Fluxes & Forces \\ \hline
  Fluid -- traceless & $\Pi_{\langle i j \rangle }$ &
    $-(\frac{1}{T} \partial_i v_j + \rho m_2 \dot \Pi_{\langle i j \rangle })$ \\ \hline
  Fluid -- spherical & $\Pi_{ii}$ &
    $-(\frac{1}{T} \partial_j v_j + \rho m_3 \dot \Pi_{jj})$ \\ \hline
  Thermal & $q_{i}$ &
    $-\rho m_1 \dot q_j + \partial_j b_{kk}  + \partial_i b_{\langle i j \rangle }$ \\ \hline
  Entropic -- traceless & $b_{\langle i j \rangle }$ &
    $\partial_{\langle i} q_{j\rangle }$ \\ \hline
  Entropic -- spherical & $b_{kk} - \frac{1}{T}$ & $\partial_i q_i$ \\
\end{tabular}\\
\caption{Thermodynamic fluxes and forces of NET-IV}
\label{table:fandf}
\end{table}
Assuming linear relations between thermodynamic fluxes and forces one realizes coupling between the quantities of same tensorial order in the case of isotropic materials.  {The complete isotropic coupling is given in \cite{FamEta18m}, but it is too complicated for a reasonable comparison, therefore we use the following simplified version: }
\begin{eqnarray}
-\rho m_1 \dot q_i +\frac{1}{3} \partial_i b_{kk}  + \partial_j b_{\langle ji \rangle }&=& n q_i, \\
-\frac{1}{T} \partial_{\langle i} v_{j \rangle }- \rho m_2 \dot \Pi_{\langle ij \rangle}  &=& l_{11}\Pi_{\langle ij \rangle} + l_{12} \partial_{\langle i} q_{j \rangle }, \\
b_{\langle i j \rangle } &=&l_{21} \Pi_{\langle ij \rangle} + l_{22} \partial_{\langle i} q_{j \rangle }, \\
- \frac{1}{T} \partial_j v_j-\rho m_3 \dot \Pi_{ii} &=& k_{11} \frac{\Pi_{ii} }{3} + k_{12} \partial_i q_i, \\
 b_{kk} - \frac{1}{T} &=& k_{21} \frac{\Pi_{ii} }{3} + k_{22} \partial_i q_i.
\end{eqnarray}

It has to be emphasized that isotropy is interpreted here in a restricted sense, including spatial reflections \cite{Mul85b}. Also, we did not assume any symmetry of the reciprocity relations. A similar assumption turned out to be extremely fruitful in the case of internal variables \cite{BerVan17b}. Positive definiteness of the entropy production constraints the sign of the coefficients:
\begin{align}
n>0, \quad l_{11} l_{22} - (l_{12}+ l_{21})^2/4 >0, \quad  k_{11} k_{22} - (k_{12} +k_{21})^2/4>0.
\end{align}

 {
From fluxes-forces relations one may eliminate the current multipliers and formally obtain the constitutive/governing equations for internal variables:
\begin{align}\label{NETSYSFINAL3Dparabolic}
  & \frac{\rho m_{1}}{n} \dot q_i + q_i + \frac{1}{3 n T^{2}} \partial_i T
    - \frac{1}{n} \partial_j \left\{ \frac{k_{21}}{9} \Pi_{kk} \delta_{ji}
    + l_{21} \Pi_{\langle ij \rangle} \right\}
  \nonumber \\
  & \quad = \frac{1}{3 n} \partial_{i} \left( k_{22} \partial_{k} q_{k} \right)
    + \frac{1}{n} \partial_{j} \left( \partial_{\langle j} q_{i \rangle} \right),
  \nonumber \\
  & \frac{\rho m_{2}}{l_{11}} \dot \Pi_{\langle ij \rangle} + \Pi_{\langle ij \rangle}
    + \frac{1}{T l_{11}} \partial_{\langle i} v_{j \rangle}
    + \frac{l_{12}}{l_{11}} \partial_{\langle i} q_{j \rangle} = 0,
  \\
  & \frac{3 \rho m_{3}}{k_{11}} \dot \Pi_{ii} + \Pi_{ii} + \frac{3}{T k_{11}} \partial_i v_i
    + \frac{3 k_{12}}{k_{11}} \partial_i q_i = 0.
  \nonumber
\end{align}
In \eqref{NETSYSFINAL3Dparabolic} we assumed that phenomenological coefficients are functions of state variables.
}

If the attention is restricted to strictly linear theory \cite{Gya77a}, that is coefficients are assumed to be constant,  {equations \eqref{NETSYSFINAL3Dparabolic} can be further simplified}:
\begin{align}\label{NETSYSLINFINAL3Dparabolic}
\tau_1\dot q_i +q_i +\lambda \partial_i T - \alpha_{21} \partial_i \Pi_{kk} - \beta_{21} \partial_j \Pi_{\langle ij \rangle} & = \gamma_{1} \partial_{i} \left( \partial_{k} q_{k} \right)
+ \gamma_{2} \partial_{j} \left( \partial_{\langle j} q_{i \rangle} \right), \nonumber \\
\tau_2 \dot \Pi_{\langle ij \rangle} +\Pi_{\langle ij \rangle} + \nu \partial_{\langle i} v_{j \rangle} + \beta_{12} \partial_{\langle i} q_{j \rangle} &= 0, \\
\tau_3 \dot \Pi_{ii} +\Pi_{ii} + \eta \partial_i v_i + \alpha_{12} \partial_i q_i &= 0, \nonumber
\end{align}
Coefficients have the following meaning: $\tau_m$ ($m=1,2,3$) are the relaxation times, $\lambda$ is the heat conductivity, $\eta$ is the bulk viscosity, $\nu$ denotes the shear viscosity, $\alpha_{ab}$, $\beta_{ab}$ ($a,b=1,2$) are the coupling parameters between the heat flux and the pressure and $\gamma_{c}$ ($c = 1,2$) are the coefficients of second order terms. These coefficients are built as follows:
\begin{gather*}
\tau_1 = \frac{\rho m_1}{n}, \quad \tau_2 = \frac{\rho m_2}{l_{11}}, \quad \tau_3=\frac{3 \rho m_3}{k_{11}}, \\
\lambda=\frac{1}{3 n T^2}, \quad \nu=\frac{1}{T l_{11}}, \eta=\frac{3}{T k_{11}}, \\
\alpha_{12}=\frac{3 k_{12}}{k_{11}} \quad \alpha_{21}=\frac{k_{21}}{9n}, \quad \beta_{12}=\frac{l_{12}}{l_{11}}, \quad \beta_{21}=\frac{l_{21}}{n}, \\
\gamma_{1} = \frac{k_{22}}{3 n}, \quad \gamma_{2} = \frac{l_{22}}{n}.
\end{gather*}

The system \eqref{NETSYSLINFINAL3Dparabolic} is parabolic due to second order contributions in \eqref{NETSYSLINFINAL3Dparabolic}$_{1}$. In order to obtain the compatibility with RET, whose equations are hyperbolic by construction, let us consider the case $\gamma_{1} = \gamma_{2} = 0$, i.e. $l_{22} = k_{22} =0$. Its consequence is that the reciprocity is antisymmetric $l_{12}=-l_{21}$ and $k_{12} =- k_{21}$. This way, one can simplify the resulting parabolic system to a hyperbolic one:
\begin{align}\label{NETSYSLINFINAL3D}
\tau_1\dot q_i +q_i +\lambda \partial_i T - \alpha_{21} \partial_i \Pi_{kk} - \beta_{21} \partial_i \Pi_{\langle ij \rangle} &= 0, \nonumber \\
\tau_2 \dot \Pi_{\langle ij \rangle} +\Pi_{\langle ij \rangle} + \nu \partial_{\langle i} v_{j \rangle} + \beta_{12} \partial_{\langle i} q_{j \rangle} &= 0, \\
\tau_3 \dot \Pi_{ii} +\Pi_{ii} + \eta \partial_i v_i + \alpha_{12} \partial_i q_i &= 0. \nonumber
\end{align}
Physical significance of the reduction from equations \eqref{NETSYSLINFINAL3Dparabolic} to the system \eqref{NETSYSLINFINAL3D} by neglecting the second order derivatives of field variables will be discussed in Section \ref{subs:10}.

\section{The 14 moments model of Rational Extended Thermodynamics}
\label{sec:14moments}

Rational extended thermodynamics of polyatomic gases with 14 moments \cite{ETdense,RETpoly} presents a different route to the description of viscous, heat-conducting gases. It fits within the general framework of RET and inherits the basic principles of modern continuum theories. It will be described briefly in the sequel for the purpose of comparison with the results of NET-IV.

\subsection{Formal structure of RET}
\label{subs:6}
There are four basic principles in this approach \cite{MulRug98b}:
\begin{itemize}
  \item[(a)] governing equations are of balance type;
  \item[(b)] constitutive relations are of local type;
  \item[(c)] governing equations are invariant with respect to Galilean transformations;
  \item[(d)] governing equations are compatible with the entropy inequality with convex entropy.
\end{itemize}
Assumptions (a) and (b) are the specific mathematical requirements introduced with the aim to obtain the system of governing equations in the form of a hyperbolic system of balance laws. Moreover, they allow the existence of weak solutions and shocks. Assumptions (c) and (d) are basic physical requirements, fully consistent with modern continuum theories. In the sequel, the basic structure of RET will be described following \cite{MulRug98b}.

Since the equations of RET have the structure of balance laws, they can be recast into the following form:
\begin{equation}\label{eq:BLaws}
  \partial_{t} \mathbf{F}^{0}(\mathbf{u}) + \partial_{i} \mathbf{F}^{i}(\mathbf{u}) = \mathbf{f}(\mathbf{u}), \quad i = 1,2,3.
\end{equation}
In \eqref{eq:BLaws} $\mathbf{u} \in \mathbb{R}^{N}$ denotes the vector of state variables, $\mathbf{F}^{0}(\mathbf{u}) \in \mathbb{R}^{N}$ is the vector of densities\footnote{It is quite common to adopt $\mathbf{F}^{0}(\mathbf{u}) \equiv \mathbf{u}$.}, $\mathbf{F}^{i}(\mathbf{u}) \in \mathbb{R}^{N}$ are the components of fluxes, and $\mathbf{f}(\mathbf{u}) \in \mathbb{R}^{N}$ is the vector of source (production) terms. Certain components of $\mathbf{F}^{0}(\mathbf{u})$ and $\mathbf{F}^{i}(\mathbf{u})$ are determined by the physical laws, but some of the fluxes are constitutive quantities, as well as the source terms $\mathbf{f}(\mathbf{u})$. Since the dependence of the densities and fluxes on $\mathbf{u}$ is of local type (in accordance with assumption (b)), the system turns into a quasi-linear system of PDE's:
\begin{equation}\label{eq:BLaws-QL}
  \mathbf{A}^{0}(\mathbf{u}) \partial_{t} \mathbf{u} + \mathbf{A}^{i}(\mathbf{u}) \partial_{i} \mathbf{u} = \mathbf{f}(\mathbf{u}),
\end{equation}
where $\mathbf{A}^{0} = \partial \mathbf{F}^{0}/\partial \mathbf{u}$ and $\mathbf{A}^{i} = \partial \mathbf{F}^{i}/\partial \mathbf{u}$. If all the eigenvalues $\lambda(\mathbf{u})$ of the eigenvalue problem $(- \lambda \mathbf{A}^{0} + n_{i} \mathbf{A}^{i}) \mathbf{d} = \mathbf{0}$ are real, where $n_{i}$ are the components of the unit-normal vector, the system \eqref{eq:BLaws-QL} is said to be hyperbolic in $t-$direction.

For any thermodynamic process described by \eqref{eq:BLaws}, in accordance with assumption (d) the state variables $\mathbf{u}$ must satisfy the entropy inequality:
\begin{equation}\label{eq:EntropyIneq}
  \partial_{t} h^{0}(\mathbf{u}) + \partial_{i} h^{i}(\mathbf{u}) = \Sigma(\mathbf{u}) \geq 0,
\end{equation}
where $h^{0}$ is the entropy density, $h^{i}$ are the components of the entropy flux, and $\Sigma$ is the entropy production rate. Due to local dependence, the entropy inequality can be put into quasi-linear form:
\begin{equation}\label{eq:Entropy-QL}
  \frac{\partial h^{0}}{\partial \mathbf{u}} \partial_{t} \mathbf{u}
    + \frac{\partial h^{i}}{\partial \mathbf{u}} \partial_{i} \mathbf{u} = \Sigma(\mathbf{u}).
\end{equation}

The entropy principle states \cite{Liu} that for any thermodynamic process the state variables $\mathbf{u}$ must obey the entropy law \eqref{eq:EntropyIneq}, provided they satisfy the balance laws \eqref{eq:BLaws}. In other words, the balance laws are regarded as \textit{constraints}. Since both the balance laws \eqref{eq:BLaws-QL} and the entropy law \eqref{eq:Entropy-QL} are quasi-linear with respect to $\mathbf{u}$, the problem with constraints can be transformed into a problem without constraints at the expense of introduction of Lagrange multipliers $\mathbf{u}^{\prime} \in \mathbb{R}^{N}$:
\begin{equation}\label{eq:EntropyLagrange}
  \partial_{t} h^{0} + \partial_{i} h^{i} - \Sigma = \mathbf{u}^{\prime} \cdot \left( \partial_{t} \mathbf{F}^{0} + \partial_{i} \mathbf{F}^{i} - \mathbf{f} \right).
\end{equation}
As a consequence, the following relations must hold:
\begin{equation}\label{eq:EntropyCompatibilityDiff}
  \mathrm{d}h^{0} = \mathbf{u}^{\prime} \cdot \mathrm{d}\mathbf{F}^{0}, \quad
    \mathrm{d}h^{i} = \mathbf{u}^{\prime} \cdot \mathrm{d}\mathbf{F}^{i},
\end{equation}
or equivalently:
\begin{equation}\label{eq:EntropyCompatibility}
  \frac{\partial h^{0}}{\partial \mathbf{u}} = \mathbf{u}^{\prime} \cdot \frac{\partial \mathbf{F}^{0}}{\partial \mathbf{u}}, \quad
    \frac{\partial h^{i}}{\partial \mathbf{u}} = \mathbf{u}^{\prime} \cdot \frac{\partial \mathbf{F}^{i}}{\partial \mathbf{u}}.
\end{equation}
Moreover, the residual inequality has to be satisfied as well:
\begin{equation}\label{eq:EntropyResidual}
  \Sigma = \mathbf{u}^{\prime} \cdot \mathbf{f} \geq 0.
\end{equation}

Equations \eqref{eq:EntropyLagrange}-\eqref{eq:EntropyResidual} require some comments. Since certain components of the fluxes are constitutive quantities, equations \eqref{eq:EntropyCompatibility} can serve for the determination of the multipliers $\mathbf{u}^{\prime}$, as well as constitutive components fluxes. Moreover, one of the legacies of RET is the assumption that entropy flux $h^{i}$ is not necessarily proportional to the heat flux, i.e., it is not defined as $h^{i} = q_{i}/T$. Instead, it is treated as a constitutive quantity, which brings certain freedom in its derivation. Finally, once the multipliers $\mathbf{u}^{\prime}$ are determined, one may choose the source terms $\mathbf{f}(\mathbf{u})$ in such a way that the residual inequality \eqref{eq:EntropyResidual} is satisfied for all thermodynamic processes. Although this statement introduces the flavor of arbitrariness, it is usually reduced to similar arguments as those used in classical TIP, or as in NET-IV in the previous Section. Namely, the components of the source terms $\mathbf{f}$ are chosen as linear forms of $\mathbf{u}^{\prime}$ (in the appropriate way which takes into account their tensorial order), so that $\Sigma$ becomes quadratic form in $\mathbf{u}^{\prime}$.

Another property of the method of multipliers---possibility to transform the system \eqref{eq:BLaws} into symmetric hyperbolic form---is substantial for the mathematical analysis, but it will not be studied here. The details, including an account on its historical development, may be found in \cite{RETpoly} (pp. 39--40).

\subsection{Hierarchical structure of governing equations}
\label{subs:7}
One of the basic properties of the governing equations of RET is that they have a hierarchical structure with increasing tensorial order of densities, fluxes, and production terms. The equations for \emph{monatomic} gas can be put into a single hierarchy, consisted of generic balance laws:
\begin{equation} \label{eq:Hierarchy1}
  \partial_{t} F_{k_{1} \cdots k_{n}} + \partial_{i} F_{k_{1} \cdots k_{n} i} = P_{k_{1} \cdots k_{n}}.
\end{equation}
The system \eqref{eq:Hierarchy1} is fully compatible with Grad's moment equations obtained in the kinetic theory of gases \cite{Grad}. However, in contrast to Grad's approach, which is relied on the approximation of non-equilibrium distribution function, closure procedure in RET is purely macroscopic and relies on the application of entropy principle and exploitation of the Liu's method of multipliers \cite{Liu}. This approach facilitated the introduction of the concept of subsystems \cite{B-R_Subsystems} with nice structural properties: Lagrange multipliers used in the exploitation of the entropy principle could be used as new state variables, so-called \textit{main field}\footnote{In mathematical community these state variables are often called \textit{entropic variables}.} \cite{R-S_Main}, and the system is transformed into a symmetric hyperbolic form.

{Since formulation of RET of monatomic gases, there were several attempts to derive the equations for 14 moments with the intention to capture the behavior of polyatomic gases. A brief review of these attempts may be found in \cite{RETpoly} (pp. 109--110). Recently \cite{ETdense}, a natural and physically convincing hierarchical structure of equations for dense and polyatomic gases was discovered. The main feature of this system is that a single-hierarchy structure of \eqref{eq:Hierarchy1} is replaced with double-hierarchy structure, which in the case of 14 moments reads:
\begin{alignat}{2}\label{eq:Hierarchy2}
  \partial_{t} F + \partial_{k} F_{k} & = 0, & \quad & \quad
    \nonumber \\
  \partial_{t} F_{i} + \partial_{k} F_{ik} & = 0, & \quad & \quad
    \nonumber \\
  \partial_{t} F_{ij} + \partial_{k} F_{ijk} & = P_{ij}, & \quad \partial_{t} G_{ll} + \partial_{k} G_{llk} & = 0, \\
  & \quad & \quad \partial_{t} G_{lli} + \partial_{k} G_{llik} & = Q_{lli}, \nonumber
\end{alignat}
where the densities may be expressed in terms of physical variables as follows:
\begin{alignat*}{2}
  F & = \rho, & \quad & \quad \\
  F_{i} & = \rho v_{i}, & \quad & \quad \\
  F_{ij} & = \rho v_{i} v_{j} + (p + \Pi) \delta_{ij} + \Pi_{\langle ij \rangle}, & \quad
    G_{ll} & = \rho v^{2} + 2 \rho e, \\
  & \quad & \quad G_{lli} & = \rho v^{2} v_{i} + 2 (\rho e + p + \Pi) v_{i} \\
  & \quad & \quad & \quad + 2 \Pi_{\langle ki \rangle} v_{k} + 2 q_{i}.
\end{alignat*}
The main feature of the hierarchy \eqref{eq:Hierarchy2} is lack of relation between energy density and momentum flux, $G_{ll} \neq F_{kk}$, i.e. $2 \rho e \neq 3 p$. Thus, dynamic pressure $\Pi$ naturally appears as the fourteenth state variable. It is convenient to identify the $F$-hierarchy as the \textit{momentum hierarchy} and $G$-hierarchy as the \textit{energy hierarchy}, according to leading non-zero order moments that appear in them.

As it was mentioned above, the hierarchical structure of RET equations for monatomic gases is compatible with Grad's moment equations. The case of polyatomic gases is more delicate due to internal degrees of freedom of molecules. Nevertheless, it was shown \cite{MEP-14} that double hierarchy \eqref{eq:Hierarchy2} can be recovered within the framework of kinetic theory (by the application of maximum entropy principle) with a single scalar variable capturing all the internal degrees of freedom (for the origin of this approach one have to consult \cite{Borg-Lars,BDTP-Poly}). This result was further developed in \cite{Arietal14} and summarized in \cite{RETpoly}.

\subsection{Constitutive relations  {for rarefied gases} near equilibrium}
\label{subs:8}
Except in rare situations, the closure problem in RET cannot be generally solved. Its solution is usually restricted to a neighborhood of a local equilibrium state. Since the application of the entropy principle determines the constitutive relations, it is common to assume the following general form of entropy density and entropy flux:
\begin{equation}\label{eq:EntropyDef}
  h^{0} = \rho s = \rho s_{eq} + \rho k, \quad
  h^{i} = h^{0} v_{i} + J_{i}
\end{equation}
where $s_{eq}(\rho, e)$ is the specific entropy in equilibrium; $k(\rho, e, \Pi, \Pi_{\langle ij \rangle}, q_{i})$ is the non-equilibrium part of the entropy density, and $J_{i}(\rho, e, \Pi, \Pi_{\langle ij \rangle}, q_{i})$ is the non-convective part of the entropy flux; they both must obey the equilibrium conditions, $k(\rho, e, 0, 0_{\langle ij \rangle}, 0_{i}) = 0$, $J_{i}(\rho, e, 0, 0_{\langle ij \rangle}, 0_{i}) = 0^{i}$.

 {
It is shown \cite{ETdense,RETpoly} that in the case of 14 moments equations, Lagrange multipliers are completely determined by the non-equilibrium specific entropy density $k$. The results of these detailed studies will be summarized in the sequel to facilitate the comparison of ET with the method of internal variables. Furthermore, the most general results of ET cover the behavior of dense gases. Here we shall restrict the analysis to the case of rarefied gases, with usual thermal and caloric equations of state:
\begin{equation}\label{eq:EqsOfState}
  p = \rho \frac{k_{\mathrm{B}}}{m} T, \quad e = e(T),
\end{equation}
where specific heat depends solely on temperature $c_{v} = c_{v}(T)$.
}

The approximate form (up to second order in non-equilibrium variables) of the entropy density and entropy flux reads:
 {
\begin{align}\label{eq:Entropy-Final}
  h^{0} & = \rho s_{eq} - \frac{3 \hat{c}_{v}}{2 (2 \hat{c}_{v} - 3)pT} \Pi^{2}
    - \frac{1}{4 \rho T} \Pi_{\langle ij \rangle} \Pi_{\langle ij \rangle}
    - \frac{\rho}{2 p^{2} T (1 + \hat{c}_{v})} q_{i} q_{i} + O(3), \\
  J_{k} & = \frac{1}{T} q_{k} - \frac{1}{p T (1 + \hat{c}_{v})} \Pi q_{k}
    - \frac{1}{p T (1 + \hat{c}_{v})} q_{i} \Pi_{\langle ik \rangle} + O(3),
  \nonumber
\end{align}
where:
\begin{equation}\label{eq:Gamma}
  \hat{c}_{v} = \frac{c_{v}}{k_{\mathrm{B}}/m},
\end{equation}
and $O(N)$ denotes the lowest order non-equilibrium terms that are neglected. Convexity arguments yield the following inequalities:
\begin{equation}\label{eq:Ineq1}
  2 \hat{c}_{v} - 3 > 0,
\end{equation}
which is always satisfied for polyatomic gases. The case of monatomic limit is carefully analyzed in \cite{RETpoly}. The source terms are expressed as linear forms of non-equilibrium variables:
\begin{equation}\label{eq:Sources-Approx}
  \hat{P}_{ll} = \frac{3}{\tau_{\Pi}} \Pi + O(2), \quad
  \hat{P}_{\langle ij \rangle} =- \frac{1}{\tau_{S}} \Pi_{\langle ij \rangle} + O(2), \quad
  \hat{Q}_{lli} = - \frac{2}{\tau_{q}} q_{i} + O(2).
\end{equation}
Using the source terms \eqref{eq:Sources-Approx} and Lagrange multipliers up to first order in non-equilibrium variables, the entropy production rate becomes:
\begin{equation}\label{eq:Production-Final}
  \Sigma = \frac{1}{\tau_{\Pi}} \frac{3 \hat{c}_{v}}{(2 \hat{c}_{v} - 3)pT} \Pi^{2}
    + \frac{1}{\tau_{S}} \frac{1}{2 p T} \Pi_{\langle ij \rangle} \Pi_{\langle ij \rangle}
    + \frac{1}{\tau_{q}} \frac{\rho}{p^{2} T (1 + \hat{c}_{v})} q_{i} q_{i} \geq 0
\end{equation}
whose definiteness is secured by the convexity inequality \eqref{eq:Ineq1} and the constraints on relaxation times:
\begin{equation}\label{eq:RelaxTimes-Final}
  \tau_{\Pi} > 0, \quad \tau_{S} > 0, \quad \tau_{q} > 0.
\end{equation}
Finally, the unknown non-convective fluxes have the following form:
\begin{align}\label{eq:Fluxes-Final}
  \hat{F}_{ijk} & = \frac{3}{1 + \hat{c}_{v}} q_{(i} \delta_{jk)} + O(2) \quad \Leftrightarrow
  \nonumber \\
  & \Leftrightarrow \quad \hat{F}_{llk} = \frac{5}{1 + \hat{c}_{v}} q_{k} + O(2), \;
    \hat{F}_{\langle ij \rangle k} = \frac{2}{1 + \hat{c}_{v}} q_{\langle i} \delta_{j \rangle k}
    + O(2)
  \\
  \hat{G}_{llij} & = 2 \left( e + \frac{k_{\mathrm{B}}}{m} T \right) \left[ (p + \Pi) \delta_{ij}
    + \Pi_{\langle ij \rangle} \right] + O(2),
  \nonumber
\end{align}
}

 {
One important feature of constitutive theory for the 14 moments equations in RET must be emphasized: all the constitutive functions are expressed in terms of the state variables, i.e., thermal and caloric equations of state (see pp. 120--123 in \cite{RETpoly}).} Only the relaxation times \eqref{eq:RelaxTimes-Final} remain undetermined. However, their values may be estimated using more refined theories like the kinetic theory of gases (see \cite{PolyMoment}) or experimental data \cite{Tanietal14}, since they are related to transport coefficients, i.e., viscosities and heat conductivity (see Eq. (5.62) in \cite{RETpoly}).

\subsection{14 moments equations (ET14)}
\label{subs:9}
Once the constitutive relations are determined, and the closure problem is resolved in accordance with the principles of RET, the complete set of governing (field) equations can be written. Although the field equations \eqref{eq:Hierarchy2} are given in conservative form, for the purpose of comparison with generalized NSF equations, we shall give them in the sequel in material form. The $F$-hierarchy consists of the conservation laws of mass and momentum, and the balance laws (evolution equations) for the dynamic pressure $\Pi$ and the deviatoric part of the pressure tensor $\Pi_{\langle ij \rangle}$ are:
\begin{align} \label{eq:BLaws14-MatF}
  & \dot{\rho} + \rho \partial_{k} v_{k} = 0,
  \nonumber \\
  & \rho \dot{v}_{i} + \partial_{j} \left[ (p + \Pi) \delta_{ij}
    + \Pi_{\langle ij \rangle} \right] = 0_{i},
  \nonumber \\
  &  {\dot{\Pi} + \left( \frac{2 \hat{c}_{v} - 3}{3 \hat{c}_{v}} p
    + \frac{5 \hat{c}_{v} - 3}{3 \hat{c}_{v}} \Pi \right) \partial_{k} v_{k}
    + \frac{2 \hat{c}_{v} - 3}{3 \hat{c}_{v}} \Pi_{\langle ik \rangle}
    \partial_{\langle i} v_{k \rangle}
  }
  \\
  & \quad  {- \frac{5}{3} \frac{1}{(1 + \hat{c}_{v})^{2}} \frac{d\hat{c}_{v}}{dT} q_{k} \partial_{k} T
    + \frac{2 \hat{c}_{v} - 3}{3 \hat{c}_{v}(1 + \hat{c}_{v})} \partial_{k} q_{k}
    = - \frac{1}{\tau_{\Pi}} \Pi,
  }
  \nonumber \\
  &  {\dot{\Pi}_{\langle ij \rangle} + \Pi_{\langle ij \rangle} \partial_{k} v_{k}
    + 2 \partial_{k} v_{\langle i} \Pi_{\langle j \rangle k \rangle}
    + 2 (p + \Pi) \partial_{\langle i} v_{j \rangle}
  }
  \nonumber \\
  & \quad  {- \frac{2}{(1 + \hat{c}_{v})^{2}} \frac{d\hat{c}_{v}}{dT}
    \partial_{k}T q_{\langle i} \delta_{j \rangle k}
    + \frac{2}{1 + \hat{c}_{v}} \partial_{\langle j} q_{i \rangle}
    = - \frac{1}{\tau_{S}} \Pi_{\langle ij \rangle}.
  }
  \nonumber
\end{align}
The $G$-hierarchy consists of the conservation law of energy, and the balance law (evolution equation) for the heat flux $q_{i}$:
\begin{align}\label{eq:BLaws14-MatG}
  & \rho \dot{e}+
    + \left[ (p + \Pi) \delta_{ij} + \Pi_{\langle ij \rangle} \right] \partial_{i} v_{j}
    + \partial_{i} q_{i} = 0,
  \nonumber \\
  &  {\dot{q}_{i} + \frac{2 + \hat{c}_{v}}{1 + \hat{c}_{v}} q_{i} \partial_{k} v_{k}
    + \frac{1}{1 + \hat{c}_{v}} q_{k} \partial_{i} v_{k}
    + \frac{2 + \hat{c}_{v}}{1 + \hat{c}_{v}} q_{k} \partial_{k} v_{i}
    }
  \\
  &  {\quad - \frac{k_{\mathrm{B}}}{m} T \partial_{i}p + \frac{k_{\mathrm{B}}}{m} \left\{
    (1 + \hat{c}_{v}) p \delta_{ki} + (2 + \hat{c}_{v})
    (\Pi \delta_{ki} + \Pi_{\langle ki \rangle}) \right\} \partial_{k}T
    }
  \nonumber \\
  &  {\quad + \frac{1}{\rho} \left\{ (p - \Pi)\delta_{ki} - \Pi_{\langle ki \rangle} \right\}
    \partial_{l} \left\{ (p + \Pi)\delta_{kl} + \Pi_{\langle kl \rangle} \right\}
    = - \frac{1}{\tau_{q}} q_{i}.
    }
  \nonumber
\end{align}

Our aim is to compare ET14 field equations with generalized NSF equations obtained in Section \ref{sec:NET-IV}.  {Although their structure is substantially different, we are in a position to make a comparison at two levels. First, we may compare non-linear NET-IV equations \eqref{NETSYSFINAL3Dparabolic} with corresponding equations \eqref{eq:BLaws14-MatF}-\eqref{eq:BLaws14-MatG}. Second, further comparison may be made if we restrict the analysis to linear regimes, i.e. the small perturbations of uniform stationary states.} To that end, we shall linearize the balance laws of ET14 model in the neighborhood of stationary equilibrium state $\mathbf{u}^{(0)} = (\rho^{(0)}, v_{i}^{(0)}, T^{(0)}, 0, 0_{ij}, 0_{i}) = \mathrm{const.}$ Field variables will be regarded as perturbations of the equilibrium state, $\mathbf{u} = (\rho^{(0)} + \rho, v_{i}^{(0)} + v_{i}, T^{(0)} + T, \Pi, \Pi_{\langle ij \rangle}, q_{i})$. Linearized balance laws then read:
 {
\begin{align} \label{eq:BLaws14-MatLin}
  & \dot{q}_{i} - \frac{k_{\mathrm{B}}}{m} T^{(0)} \partial_{i}p 
    + \frac{k_{\mathrm{B}}}{m} (1 + \hat{c}_{v}^{(0)}) p^{(0)} \partial_{i}T 
  \nonumber \\
  & \quad + \frac{p^{(0)}}{\rho^{(0)}} \partial_{k} \left\{ (p + \Pi) \delta_{ik} 
    \Pi_{\langle ik \rangle} \right\} = - \frac{1}{\tau_{q}} q_{i}.
  \nonumber \\
  & \dot{\Pi}_{\langle ij \rangle} + 2 p^{(0)} \partial_{\langle j} v_{i \rangle}
    + \frac{2}{1 + \hat{c}_{v}^{(0)}} \partial_{\langle j} q_{i \rangle} 
    = - \frac{1}{\tau_{S}} \Pi_{\langle ij \rangle},
  \\
  & \dot{\Pi} + \frac{2 \hat{c}_{v}^{(0)} - 3}{3 \hat{c}_{v}^{(0)}} p^{(0)} \partial_{k} v_{k}
    + \frac{2 \hat{c}_{v}^{(0)} - 3}{3 \hat{c}_{v}^{(0)} (1 + \hat{c}_{v}^{(0)})} \partial_{k} q_{k}
    = - \frac{1}{\tau_{\Pi}} \Pi.
  \nonumber
\end{align}}
We did not linearize the conservation laws since they are completely equivalent. Note that we ordered linearized balance laws \eqref{eq:BLaws14-MatLin} in the same way as in \eqref{NETSYSLINFINAL3D} for the sake of easier comparison.

\section{Comparison and summary}
\label{sec:Comparison}

In this Section, we shall compare the equations, the procedure of their derivation, and the underlying assumptions of two approaches in the modeling of rarefied gases. First, the governing equations will be compared at formal level. After that, two approaches will be compared for the concepts and procedures used in the study of non-equilibrium rarefied gases.

\subsection{Comparison of the governing equations}
\label{subs:10}
Governing equations of the NET-IV model consist of conservation laws of mass, momentum, and energy \eqref{IV-BLaws}$_{1-3}$ and constitutive relations \eqref{NETSYSFINAL3Dparabolic}. The latter has the form of balance laws describing the time rate of change of heat flux $q_{i}$, pressure deviator $\Pi_{\langle ij \rangle}$ and dynamic pressure $\Pi_{ii}$. On the other hand, the ET14 model is structured in two hierarchies. $F-$hierarchy \eqref{eq:BLaws14-MatF} consists of the conservation laws mass and momentum, and the balance laws for dynamic pressure $\Pi = (1/3)\Pi_{ii}$ and stress deviator $\Pi_{\langle ij \rangle}$; $G-$hierarchy \eqref{eq:BLaws14-MatG} consists of the conservation law of energy and the balance law for heat flux $q_{i}$.

Conservation laws in both models are completely equivalent. The difference appears in the structure of balance laws.  {NET-IV balance laws \eqref{NETSYSFINAL3Dparabolic} are derived by linear coupling of thermodynamic forces and fluxes, while balance laws in ET14 model are quasi-linear by assumption, with state-dependent coefficients.} To make all the formal differences as clear as possible, we shall list them below:
\begin{itemize}
\item[(a)] balance law \eqref{NETSYSFINAL3Dparabolic}$_{1}$ contains second order parabolic terms in $q_{i}$, which are not present in \eqref{eq:BLaws14-MatG}$_{2}$;
\item[(b)]  {balance law \eqref{eq:BLaws14-MatG}$_{2}$ is apparently coupled with velocity gradient field through $\partial_{k} v_{k}$ and $\partial_{k} v_{i}$; this coupling does not appear in \eqref{NETSYSFINAL3Dparabolic}$_{1}$; other couplings with gradients of state variables are implicit in \eqref{NETSYSFINAL3Dparabolic}$_{1}$, since phenomenological coefficients may depend on them;}
\item[(c)]  {balance law \eqref{eq:BLaws14-MatF}$_{4}$ has rather complex quasi-linear form; similar coupling is present in \eqref{NETSYSFINAL3Dparabolic}$_{2}$ only in part where $\partial_{\langle i} v_{j \rangle}$ and $\partial_{\langle i} q_{j \rangle}$ appear;}
\item[(d)]  {finally, balance law \eqref{eq:BLaws14-MatF}$_{3}$ is coupled with deviatoric part of velocity gradient $\partial_{k} v_{j}$, as well as gradient $T$, while they do not appear in \eqref{NETSYSFINAL3Dparabolic}$_{3}$.}
\end{itemize}
 {Common difference between ET14 balance laws and NET-IV constitutive relations lies in the presence of quasi-linear coupling between non-equilibrium (internal) variables $\Pi$, $\Pi_{\langle ij \rangle}$ and $q_{i}$ with gradients of state variables $\partial_{i} \rho$, $\partial_{k} v_{k}$, $\partial_{\langle j} v_{i \rangle}$ and $\partial_{i} T$, and the presence of second order parabolic terms. For example, terms like $\Pi \partial_{k} v_{k}$, $\Pi_{\langle ik \rangle} \partial_{\langle i} v_{k \rangle}$, $q_{k} \partial_{k} T$ and similar ones, present in \eqref{eq:BLaws14-MatF} and \eqref{eq:BLaws14-MatG}, do not appear at all in constitutive relations \eqref{NETSYSFINAL3Dparabolic}. On the other hand, NET-IV equations are parabolic by construction, whereas ET14 equations get parabolic terms through closure procedure. These facts lead to a conclusion that both models capture non-equilibrium effects, albeit in a different way---ET14 equations through quasi-linear coupling and NET-IV equations through parabolic terms. This could also motivate the application of ET14 equations to the analysis of processes in which rapid changes of state variables occur, while NET-IV equations go deeper into the processes with prominent changes in internal variables (in this case $q_{i}$) and moderate variation of state variables.}

 {Further comparison of the NET-IV and RET systems depends on the level of the introduced simplifications. For example, it is easy to compare} linearized version \eqref{NETSYSLINFINAL3Dparabolic} of constitutive relations with constant phenomenological coefficients with the linearized form of balance laws \eqref{eq:BLaws14-MatLin}. As we already anticipated in Section \ref{sec:NET-IV} and \ref{subs:9}, further simplifications ought to be done:}
\begin{itemize}
  \item[(1)] balance laws \eqref{NETSYSLINFINAL3Dparabolic} of NET-IV model have to be reduced to hyperbolic form; therefore, it has to be assumed $\gamma_{1} = \gamma_{2} = 0$ which reduces the system to \eqref{NETSYSLINFINAL3D};
  \item[(2)] balance laws of the ET14 model have to be linearized; linearization in the neighborhood of constant equilibrium state leads to the balance laws in the form \eqref{eq:BLaws14-MatLin}.
\end{itemize}
A simple comparison of the reduced systems of balance laws leads to the following identification of the coefficients:
 {
\begin{gather} \label{eq:Comparison-Coeff}
  \tau_{1} = \tau_{q}^{(0)}, \quad \tau_{2} = \tau_{S}^{(0)}, \quad \tau_{3} = \tau_{\Pi}^{(0)},
  \nonumber \\
  \lambda = \frac{k_{\mathrm{B}}}{m} (1 + \hat{c}_{v}^{(0)}) p^{(0)} \tau_{q}^{(0)}, \quad
    \nu = 2 p^{(0)} \tau_{S}^{(0)}, \quad 
    \eta = \frac{2 \hat{c}_{v}^{(0)} - 3}{\hat{c}_{v}^{(0)}} p^{(0)} \tau_{\Pi}^{(0)},
  \nonumber \\
  \alpha_{12} = \frac{2 \hat{c}_{v}^{(0)} - 3}{\hat{c}_{v}^{(0)}(1 + \hat{c}_{v}^{(0)})}
    \tau_{\Pi}^{(0)}, \quad
    \alpha_{21} = - \frac{1}{3} \frac{p^{(0)}}{\rho^{(0)}} \tau_{q}^{(0)},
  \\
  \beta_{12} = \frac{2}{1 + \hat{c}_{v}^{(0)}} \tau_{S}^{(0)}, \quad 
    \beta_{21} = - \frac{p^{(0)}}{\rho^{(0)}} \tau_{q}^{(0)}.
  \nonumber
\end{gather}}
Note that this comparison leads to equivalent results for transport coefficients as thermodynamic limit analysis in RET \cite{ETdense,RETpoly}.

The comparative analysis given above revealed common points of the governing equations, as well as their differences. We shall close this analysis with two remarks concerned with the physical background of formal comparison.

First, we may note that second-order parabolic terms, that appear in \eqref{NETSYSLINFINAL3Dparabolic}$_{1}$, look like a higher-order regularization of the hyperbolic balance law. To that end, a comparison with the so-called R13 (regularized 13 moments) equations could be made. Their proper form, restricted to linear regimes, may be found in \cite{StruchtTorr13}, equations (2)-(3). Formal comparison immediately yields that \eqref{NETSYSLINFINAL3Dparabolic}$_{1}$ is fairly similar to equation (2) from \cite{StruchtTorr13} -- the only missing term is the derivative of deviatoric part of the velocity gradient. Nevertheless, the remaining higher-order terms are present, and represent the third-order terms in Knudsen number. Therefore, parabolic terms in \eqref{NETSYSLINFINAL3Dparabolic}$_{1}$ tend to capture heat conduction effect in rarefied gases in the so-called transition regime. When the Knudsen number is sufficiently small, we may neglect these terms by taking $\gamma_{1} = \gamma_{2} = 0$. Note, however, that no such terms appear in \eqref{NETSYSLINFINAL3Dparabolic}$_{2-3}$, which is a consequence of the closure procedure that we applied.

The second remark is concerned with a possible extension of comparative analysis in the way which puts Meixner theory into a common perspective with NET-IV and ET14 models. It was already shown in Section \ref{subs:5} that IV-Meixner model \eqref{IV-BLaws} can be reduced to form \eqref{IV-BLawsReduced} compatible with \eqref{ET6-BLawsReduced} by neglecting pressure deviator ($\mu = 0$) and heat flux ($\lambda = 0$). On the other hand, it is a common procedure in RET \cite{ET,RETpoly} to study classical thermodynamic limit using the so-called Maxwellian iteration---an asymptotic procedure which is akin to expansion in powers of relaxation time. It leads to the recovery of classical constitutive relations for pressure tensor and heat flux. Its application to ET14 equations leads to the NSF model with dynamic pressure and bulk viscosity. However, if we take into account that relaxation time for dynamic pressure is several orders of magnitude greater than relaxation times for pressure deviator and heat flux (see \cite{Tanietal14}), it seems reasonable to apply Maxwellian iteration to the ET14 system \eqref{eq:BLaws14-MatF}-\eqref{eq:BLaws14-MatG} partially, i.e. to apply it to linearized balance laws \eqref{eq:BLaws14-MatLin}$_{1}$ for heat flux and \eqref{eq:BLaws14-MatLin}$_{2}$ for pressure deviator and reduce them to constitutive relations \eqref{IV-BLaws}$_{5,6}$, but to retain the balance law \eqref{eq:BLaws14-MatLin}$_{3}$ intact. Assuming $l_{22} = 0$, almost complete compatibility is reached with \eqref{IV-BLaws}---there is only one extra term in \eqref{eq:BLaws14-MatLin}$_{3}$, $\partial_{i} q_{i}$, which does not appear in \eqref{IV-BLaws}$_{4}$.

\subsection{Comparison of concepts and procedures}
\label{subs:11}
This Section aims to analyze the concepts which stand behind the NET-IV and RET. Since concepts are closely related to the procedures applied for the derivation of equations, they will be compared as well.

The formalism of NET-IV is based upon the following premises:
\begin{itemize}
  \item[(IV-a)] The governing equations are standard balances of mass, momentum, and energy, with undetermined non-convective fluxes and by the evolution equations of internal variables. The standard balances are constraints in the calculated entropy production, the evolution equations of the internal variables are constitutive.
  \item[(IV-b)] State variables are the densities of the conservation laws and the internal variables. Non-equilibrium effects are included through the internal variable dependent part of the entropy density, and through current multipliers, which are the constitutive elements of the entropy flux.
  \item[(IV-c)] The closure is achieved through the entropy inequality. The entropy production rate has bilinear form---as the sum of products of thermodynamic forces and fluxes. The simplest linear relation between forces and fluxes provides the evolution equations of the internal variables.
  \item[(IV-d)] Elimination of internal variables and current multipliers yield constitutive relations for physical variables, e.g., non-convective fluxes. These relations may have the form of balance laws.
\end{itemize}
On the other hand, the formalism of RET provides a different setting:
\begin{itemize}
  \item[(ET-a)] The governing equations consist of two finite hierarchies of balance laws, with  increasing tensorial order. They have a nested structure---fluxes in the equation of order $n$ become densities in the equation of order $n+1$.
  \item[(ET-b)] The state variables are the densities. Unknown fluxes and source terms are local functions of state variables.
  \item[(ET-c)] The closure problem is resolved through the compatibility of balance laws and the entropy inequality using the method of multipliers.
  \item[(ET-d)] Compatibility conditions lead to the (approximate) form of non-equilibrium entropy density and entropy flux, and (approximate) closure for the unknown fluxes and source terms, which yield the definiteness of the entropy production rate.
\end{itemize}

It is clear from the procedures described above that there are two common pillars on which both approaches stand---standard conservation laws and the entropy balance law. The differences emerge in the strategy chosen to capture the non-equilibrium phenomena. NET-IV starts with the non-equilibrium entropy density and the entropy flux, which inherit internal variables and current multipliers and get the constitutive relations from the entropy production rate. Evolution equations of the internal variables emerge in this procedure and may have the form of balances. Instead, RET proceeds in a bit opposite direction: it starts with an extended list of state variables governed by the appropriate balance laws. By means of compatibility with entropy balance law, one determines the structure of entropy density and entropy flux, as well as the non-convective fluxes, while residual inequality serves for determination of the source terms in balance laws. It is, however, interesting that both approaches, in the end, provide equivalent entropy densities and entropy fluxes---compare Eqs. \eqref{IV-MeixnerEntropy} and \eqref{ET6-Entropy-EntropyFlux} for Meixner theory, and Eqs. \eqref{NET-IV-Entropy}, \eqref{JC} and \eqref{eq:Entropy-Final} for generalized NSF/14 moments equations.
We also compare the conceptual differences in a different way.

\begin{enumerate}

\item{\em Universality.} NET-IV is based on general assumptions. Beyond the second law, there was no assumption regarding the structure of the material. The internal variables and current multipliers are convenient representations of the deviation from local equilibrium. For example, instead of the multiplier structure, one can use the additive extension of the entropy flux with $k$ vector, but in that form, the simple linear solution of the entropy inequality is less apparent \cite{Ver83a,Nyiri91,BerVan17b}. On the other hand, the assumed hierarchical structure of the RET governing equations, and their compatibility with kinetic theory introduces formal restrictions on the structure of balance laws for the extended set of state variables.

\item{\em Balance structure.} In the case of RET, it is inherited from kinetic theory and due to the balance form of the Boltzmann equation. For NET-IV it is the consequence of the second law.

\item{\em Locality.} The locality of RET is required from the outset and leads to the quasi-linear hyperbolic structure of equations. They are compatible with the moment equations of kinetic theory as well \cite{MEP-14,PolyMoment,Arietal14}. Then the usual gradient terms in the constitutive equations emerge by the elimination of the higher moments \cite{Rug12a}. In our case, the missing Fourier term in eqs. \eqref{eq:BLaws14-MatG}$_{2}$ is remarkable.

\item{\em Spacetime embedding.} In RET, it is based on the invariance of the balance structure for Euclidean transformations and, in particular, for Galilean ones. With this starting point, the transformation rules of the particular physical quantities can be derived \cite{Rug89a,MulRug98b}. It is remarkable that Galilean relativity of fluids, that is a reference frame independent covariant treatment, leads to the same transformation rules of the physical quantities. In that treatment, the balances are natural four-divergences \cite{Van17a,VanEta19a}. The generalization of the entropy four-vector with internal variables and the role of current multipliers are not yet understood in the reference frame independent approach.

\item{\em Second law.} The assumption of quadratic entropy density is common in both theories. RET assumes a particular entropy flux, as it is required by the Lagrange multipliers. The entropy inequality is calculated in NET-IV, and the constitutive quantities, including the evolution equations of internal variables, are determined accordingly. In RET, the entropy inequality is calculated and determines the sign of the relaxation times.

Another interesting example is shown in \cite{SzuFul9a}, where further aspects are discussed in case of mechanics.

\item{\em Nonlinearity.} The apparent nonlinearity of the RET equations is mostly due to state-dependent transport coefficients. The assumed particular nonlinear thermodynamic coefficients can be calculated. We have proved that a nonlinear version of NET-IV, with state-dependent transport coefficients, is equivalent to the 6 field ET6 model. The complete identification is different from the one given in \cite{Arietal15}. It is not clear whether such identification exists for the nonlinear 14 field theories of NET-IV and RET.

\item{\em Further extension}. It is possible in both theories. However, with further extensions, the differences are seemingly increasing. See e.g., the treatment of NET-IV with the Nyíri multiplier related to the tensor internal variable. The complete treatment requires higher-order isotropic representations, and the number of coefficients is increasing, see in \cite{FamEta18m}.

\item{\em Experiments.} In case of ballistic heat conduction, with a single hierarchical structure, NET-IV seems to reproduce the data better \cite{KovVan15,KovVan18}. There is an extensive analysis of rarefied gas experiments beyond the Navier-Stokes-Fourier regime with RET \cite{RETpoly}.
\end{enumerate}

The last point in the comparison of conceptual differences also provides a strong motivation for the  further simultaneous analysis of the NET-IV and RET approach to polyatomic gases. Namely, there are numerous studies of experimental character in the literature, showing the non-classical behavior of rarefied gases. Sluijter et al. \cite{SluiEtal64,SluiEtal65} measured the sound absorption at low temperatures ($77 \mathrm{K}$ and $90 \mathrm{K}$) and at room temperature also, with different types of molecules. Similar measurements are made by Rhodes \cite{Rhod46}, Greenspan \cite{Gre56} and Sette et al. \cite{SetEtal55}. From the kinetic theory point of view, these experimental results were investigated in detail by Arima et al. \cite{Arietal12,AriEtal12c,Arietal13,Arietal15}. The next step is to compare the modeling capabilities of different theories to each other based upon the measurement data.

Another line of research will be related to a paradigmatic problem in non-equilibrium processes---the shock structure problem. It was studied in the framework of RET for 14 moments equations \cite{Tanietal14}, for ET6 model \cite{Tanietal14b,Pavicetal17} and from the point of view of the kinetic theory of gases \cite{KosugeAoki18}. There remains an open problem of the application of the NET-IV approach in this context, which will undoubtedly bring new insights.

As a summary, we want to emphasize that the two theoretical approaches must be concordant. Both second law based methods and kinetic theory should provide a mechanism to obtain universal evolution equations. In this study, we have shown that the compatibility is valid, but there are still several points of deviations where the relation of the two approaches is not clear. This comparative analysis can be helpful for future improvements.

\section{Acknowledgments}
The work was supported (R. Kov\'{a}cs and P. V\'{a}n) by the grants of National Research, Development and Innovation Fund -- NKFIH 124366(124508), NKFIH 123815, NKFIH 130378, TUDFO/51757/2019-ITM, Thematic Excellence Program and (D. Madjarevi\'{c} and S. Simi\'{c}) by the Ministry of Education, Science and Technological Development,
Republic of Serbia, within the project ``Mechanics of nonlinear and dissipative systems---contemporary models, analysis and applications'', Project No. ON174016.


%
%

\end{document}